# On-Chip Diamond Raman Laser

Pawel Latawiec, Vivek Venkataraman, Michael J. Burek, Birgit J. M. Hausmann*, Irfan Bulu** and Marko Lončar

School of Engineering and Applied Sciences, Harvard University, Cambridge, MA, USA

*Current affiliation: Department of Chemistry, UC Berkeley, and Materials Sciences Division, LBNL, Berkeley, CA

**Current affiliation: Schlumberger-Doll Research Center, Cambridge, MA, USA

**Synthetic single-crystal diamond has recently emerged as a promising platform for Raman lasers at exotic wavelengths due to its giant Raman shift, large transparency window and excellent thermal properties yielding a greatly enhanced figure-of-merit compared to conventional materials [1, 2, 3]. To date, diamond Raman lasers have been realized using bulk plates placed inside macroscopic cavities [3, 4, 5, 6, 7, 8, 9, 10, 11, 12, 13], requiring careful alignment and resulting in high threshold powers (~W-kW). Here we demonstrate an on-chip Raman laser based on fully-integrated, high quality-factor, diamond racetrack micro-resonators embedded in silica. Pumping at telecom wavelengths, we show Stokes output discretely tunable over a ~100nm bandwidth around 2-μm with output powers >250 μW, extending the functionality of diamond Raman lasers to an interesting wavelength range at the edge of the mid-infrared spectrum [14]. Continuous-wave operation with only ~85 mW pump threshold power in the feeding waveguide is demonstrated along with continuous, mode-hop-free tuning over ~7.5 GHz in a compact, integrated-optics platform.**

Diamond serves as a compelling material platform for Raman lasers operating over a wide spectrum due to its superlative Raman frequency shift (~40 THz), large Raman gain (~10 cm/GW @ ~1-μm wavelength) and ultra-wide transparency window (from UV (>220nm) all the way to THz, except for a slightly lossy window from ~2.6 – 6 μm due to multiphonon-induced absorption) [1, 15]. Furthermore, the excellent thermal properties afforded by diamond (giant thermal conductivity of ~1800 W/m/K @ 300K and low thermo-optic coefficient of ~$10^{-5}$ /K) [1, 2] along with negligible birefringence [3, 15] make it an ideal material for high-power Raman lasing with greatly reduced thermal lensing effects [1, 3].

The availability of CVD-grown, high-quality polished, single-crystal diamond plates has enabled the development of bulk Raman lasers using macroscopic optical cavities across the UV [6], visible [4, 5], near-infrared [7, 8, 9, 10, 11, 12] and even mid-infrared [13] regions of the optical spectrum. Although showing great performance with large output powers (many Watts) [12] and near quantum-limited conversion efficiencies [5, 9], most operate in pulsed mode in order to attain the very high pump powers required to exceed the Raman lasing threshold [5, 6, 11, 12]. Demonstration of continuous-wave diamond Raman lasing has been challenging, with very few reports [3, 7, 8]. Bulk cavity systems also require precise alignment and maintenance of optical components for the laser to function robustly.

Translating Raman laser technology onto an integrated optics platform where the light is confined to nano-waveguides [16, 17] and/or high quality-factor (Q) micro-resonators [18, 19, 20, 21] can greatly reduce pump power requirements and enable stable continuous-wave (CW) operation without the need for any complicated alignment of optical components. Such compact micro-resonator-based Raman lasers, especially if integrated on-chip, might be particularly useful for spectroscopy and sensing applications in harsh environments [22, 23] as well as medical device technologies [22, 24]. To date, chip-based Raman microlasers have been demonstrated in silicon racetracks [21, 25] and photonic crystals [20], and silica microtoroids [19]. Such telecom-laser-pumped devices have shown impressive performance like CW lasing with low threshold powers (µW-mWs), albeit at limited Stokes wavelengths around ~1.6 - 1.7 µm and cascaded operation out to 1.85 µm [21]. This is because of the relatively low value of the Raman frequency shift in silicon (~15.6 THz) and silica (~12.5 THz) compared to diamond (~40 THz). Moreover, the losses due to two-photon and free carrier absorption in silicon need to be mitigated via carrier extraction that complicates the device layout and fabrication process [17, 20, 21, 25]. Silica-based devices require ultrahigh Q cavities (~$10^8$) to effectively compensate for the extremely low Raman gain coefficient (>100x smaller than silicon and diamond). Additionally, the broad Raman gain spectrum in silica (~10 THz) makes single-mode operation non-trivial [18, 19]. These devices (microspheres, microtoroids) are also difficult to integrate into a compact, fully-integrated on-chip package, requiring careful alignment of a tapered fiber to evanescently couple light into the resonator [19]. Finally, both silica and silicon suffer from severe thermal management issues, and absorption losses outside of their traditional operating windows, raising a question mark on high-power operation over a wide spectrum.

Diamond can potentially overcome these drawbacks and has recently emerged as a novel nanophotonics material with applications in integrated, on-chip quantum [26, 27] and nonlinear optics [28]. Diamond's large bandgap of ~5.5 eV and lack of Reststrahlen-related absorption at low frequencies affords it a wide space for creating high quality factor resonators. Here we demonstrate the first CW, tunable, on-chip Raman laser operating at ~2-µm wavelengths using telecom-laser-pumped high-Q, waveguide-integrated diamond racetrack resonators embedded in silica on a silicon chip.

The Raman process (Fig. 1a) involves scattering of a high energy pump photon at frequency $\omega_P$, into a low energy Stokes photon at frequency $\omega_S$, via the creation of an optical phonon of frequency $\Omega_R$, such that $\omega_P - \omega_S = \Omega_R$. For diamond, $\Omega_R$ ~40 THz, corresponding to high-energy optical phonons vibrating along the <111> direction [1, 9]. For pump wavelengths in the telecom range ($\lambda_P$ ~ 1.6 µm), $\omega_P$ ~ 190 THz, resulting in a Stokes wavelength near ~2 µm ($\omega_S$ ~ 150 THz). Our diamond waveguides, with ~700x800nm cross-section embedded in silica, support modes both at the pump and Stokes wavelengths with good spatial overlap (Fig. 1b). Raman scattering does not require any phase matching as it is an inelastic process. The efficiency of this process, however, is very low in bulk materials and can be significantly increased using optical cavities. In particular, if the cavity is resonant with the Stokes wavelength it can provide optical feedback needed to stimulate the scattering process, which can lead to lasing action. If the cavity is also resonant at the pump wavelength, it can boost up the pump intensity by a factor of the finesse and further enhance the stimulated process. The threshold for Raman lasing in such a doubly-resonant cavity is inversely proportional to the product of the Qs of the pump and Stokes modes [18, 19]. The Raman gain spectrum in diamond is extremely narrow with a full-width at half-maximum (FWHM) of ~60 GHz [1, 2]. To ensure that a resonator mode exists close to the gain maximum, long racetrack micro-resonators (path length ~600 μm) are fabricated (see Methods) with free spectral range (FSR ~180 GHz) approaching the Raman scattering linewidth (Fig. 1c,d) [28, 29]. Transmission measurements (see Methods) revealed that the diamond resonators support high-Q modes at both the telecom pump (Fig. 2a) and ~2-μm Stokes wavelengths (Fig. 2b). The modes at telecom were found to be under-coupled with ~30-40% transmission dips on-resonance and high loaded Qs around 400,000 (Fig. 2a). The higher-wavelength modes around 2-μm also showed under-coupling with ~30-40% extinction ratios on-resonance and loaded Qs around 30,000, although this may have been limited by the resolution of our optical spectrum analyzer.

When the pump laser is tuned into a resonance with sufficient power, Raman lasing at the Stokes wavelength is observed (see Methods). Fig. 3a shows the measured optical spectrum with the Stokes line ~40 THz away from the pump. A zoom into the Stokes line (inset of Fig. 3a) shows resolution-limited linewidth and >60 dB sideband suppression ratio after correcting for losses, characteristic of low-noise single-mode operation. Fig. 3b shows the measured output Stokes power as a function of input pump power, displaying a clear threshold and onset of Raman lasing at ~85 mW of CW power in the coupling waveguide. Stokes powers >250 μW are coupled into the output waveguide, corresponding to an external conversion slope efficiency above threshold of ~0.43%. This is limited by the severely undercoupled nature of the resonances at both the pump and Stokes [18, 19], and the internal quantum efficiency itself is estimated to be ~12%. Knowing the Q-factor and mode volume of our device enables us to extract an effective Raman gain value of ~2.5 cm/GW from the Raman lasing threshold formula [18, 19]. This is comparable to, but lower than, previous estimates for diamond at these wavelengths (~6 cm/GW) [1], suggesting that our Stokes mode is probably not positioned exactly on the Raman gain peak. We also demonstrate discrete tuning of the Raman laser over a wide bandwidth by tuning the pump laser to separate adjacent resonances. Fig 4a shows the result of 14 separate measurements which show a Raman signal spanning from <1950 nm to >2050 nm. The discrete tuning range is >100 nm, or ~7.5 THz, which corresponds to ~5% of the center frequency and was limited by the operation bandwidth of our pump amplifiers. Within this range, over 40 uniformly spaced longitudinal modes can be individually addressed, each separated by the cavity FSR of ~180 GHz. Continuous, mode-hop free tuning of the Stokes output over ~7.5 GHz was also demonstrated (Fig. 4c) by tuning the pump within a single thermally red-shifted resonance (see Methods). In order to create a Raman laser which can be tuned over the entire output range continuously, it should suffice to create a resonator with a sufficiently small FSR on the order of the thermal shift (this would require a resonator length ~10x our current device which should be possible via a winding spiral resonator design). Then, by tuning into a mode and using its redshift (or, alternatively, an external heater), it should be possible to sweep across one resonance and carry the Stokes from one longitudinal mode of the resonator to the next continuously [21].

In conclusion, we have demonstrated a CW, low-threshold, tunable, on-chip Raman laser operating at ~2-μm wavelengths based on waveguide-integrated diamond racetrack microresonators. The threshold power is limited by the severe under-coupling of the bus waveguide to the resonator, and could be further reduced by moving to near critically-coupled modes for the pump [18, 19]. This can be easily achieved, for example, by slightly reducing the coupling-gap between the bus-waveguide and resonator. The external conversion efficiency can also be drastically increased by having over-coupled resonances for the Stokes in addition to critical-coupling for the pump [18, 19], and this should naturally happen in the current design if the intrinsic Qs of the pump and Stokes mode are of the same order. Longer coupling sections and other coupling designs can also be investigated [21]. Further improvement can be made by having higher intrinsic Q [28] and/or smaller FSR (to ensure maximum Raman gain) i.e. longer path-length resonators [21]. Another limiting factor comes from the orientation of the diamond itself. Our devices are fabricated in [100]-oriented diamond, and the pump and Stokes mode are both TE polarized, where Raman gain is sub-optimal and there is no polarization preference for the Stokes [1, 9]. By ensuring that the light polarization is parallel to <111>, for example using angle-etched resonators [30, 31] in thick [111] diamond plates, the efficiency of the Raman process can be enhanced [1, 9]. Further, by moving to an all-diamond structure, the resonator should be able to support more circulating power and reach higher output powers while also offering a route toward longer-wavelength/cascaded Raman lasers, where the absorption of silica would limit performance otherwise. Nonetheless, the current platform already offers a large amount of flexibility, with the option to fabricate devices at visible wavelengths, where the Raman gain is ~20x higher [1]. Operation in the visible would enable integration of classical nonlinear optics technologies (Raman lasing, Kerr frequency combs) with the quantum optics of color centers [26, 27, 28].

## Methods

**Device Fabrication**

The basic fabrication process was developed from the previously described approach for integrated diamond devices [26, 28, 29]. Initially, a ~20-μm thick type IIa CVD single crystal diamond (Delaware Diamond Knives) was cleaned in a refluxing acid mixture of nitric, sulphuric, and perchloric in equal ratios. The device was then thinned to specification (<1 μm) by cycling $Ar/Cl_2$ and $O_2$ etching steps in a dedicated Plasmatherm inductively-coupled-plasma reactive-ion-etcher (ICP-RIE) while bonded via Van der Waals forces to a sapphire carrier wafer [26]. The diamond was etched on both sides to remove residual stress/strain from the polishing procedure. Afterwards, the thin diamond film was transferred to a $SiO_2$/Si substrate with a 2-um thermal $SiO_2$ layer. To promote resist adhesion, a thin layer (<5 nm) of $SiO_2$ was deposited via atomic layer deposition. Then, an etch mask was patterned using Fox 16 electron-beam resist (spin-on-glass, Dow Corning) in an electron-beam lithography tool (Elionix ELS-F125) under multipass exposure. The faces of the supplied thin diamond plates are non-parallel due to the polishing process, with a thickness wedging of ~300 nm per ~1 mm length. The pattern was aligned to the diamond thin film such that the polishing gradient ran parallel to the racetrack devices. This pattern was then etched into the diamond with a final oxygen etch. The Fox 16 resist was left on the diamond. The completed waveguide had dimensions of ~800 nm in width and ~700 nm in height, while the coupling region had a gap of around ~500 nm. The diamond bus waveguide tapered off over a length of ~200 μm to an end width of ~150 nm. Polymer coupling pads to the end of the substrate were written in SU-8 aligned to the adiabatically tapered diamond waveguides [29]. Finally, a layer of ~3-μm of silica was deposited with plasma-enhanced chemical vapor deposition (PECVD) in order to cap the devices and aid in the polishing of the end facets.

**Optical measurements**

The on-chip diamond resonators are characterized using a lensed-fibre-based coupling setup [29, 28]. Transmission measurements at telecom were taken by sweeping a continuous-wave laser (Santec TSL-510) across the resonances and sending the output to an amplified photodetector (Newport 1811). The insertion loss for the device was measured to be ~5 dB per facet (~10 dB total loss from input to output lensed fibre) for telecom wavelengths. In order to measure the resonator modes around the Stokes wavelengths near 2-μm, a broadband supercontinuum source (NKT Photonics SuperK) was coupled into the device and the output spectrum was recorded on an optical spectrum analyzer (OSA, Yokogawa AQ6375) with a maximum resolution of 0.056 nm. The insertion loss for the device was measured to be ~9.5 dB per facet (~19 dB total loss from input to output lensed fibre) at these longer wavelengths. For Raman lasing measurements, high pump power was achieved by boosting the input laser power through either a C-band (~1535-1570 nm) or L-band (~1570-1610 nm) erbium-doped fiber amplifier (EDFA, Manlight). The laser was first set at a slightly blue-detuned position near a resonance before slowly being shifted into it. Power absorbed by the resonator and its host material causes a thermal redshift of the resonance resulting in a characteristic 'shark-fin' shape, allowing the pump to be slowly tuned towards the transmission minimum while stabilizing the power coupled into the resonator [21, 28]. While tuning the pump, the Stokes output was monitored on the OSA. After the onset of Raman lasing at a particular detuning, the pump was further fine-tuned to maximize the output. The continuous-tuning range shown in Fig. 4c was obtained in this manner by tuning the pump within one thermally red-shifted resonance [21]. As the intra-cavity power increased, the pump and lasing mode were both shifted to the red. Beyond the resonance (sharp edge of the 'shark-fin'), the mode was no longer pumped and began to cool down, shifting the resonance back to its original position.

## Acknowledgements

Devices were fabricated in the Center for Nanoscale Systems (CNS) at Harvard. The authors thank Dan Twitchen and Matthew Markham for helpful discussions and for diamond samples. This work was supported in part by the NSF under grant ECCS-1202157.

## Author contributions

M.L. and I.B. conceived, and together with all the other authors, designed the experiment. The theoretical studies, numerical modelling and design were carried out by V.V. and I.B. Devices were fabricated by P.L. and B.J.M.H. Experiments were performed by P.L. and V.V. with help from M.J.B. Data were analyzed by P.L and V.V. and discussed by all the authors. P.L., V.V. and

M.L. wrote the manuscript in discussion with all the authors. M.L. is the principal investigator on the project.

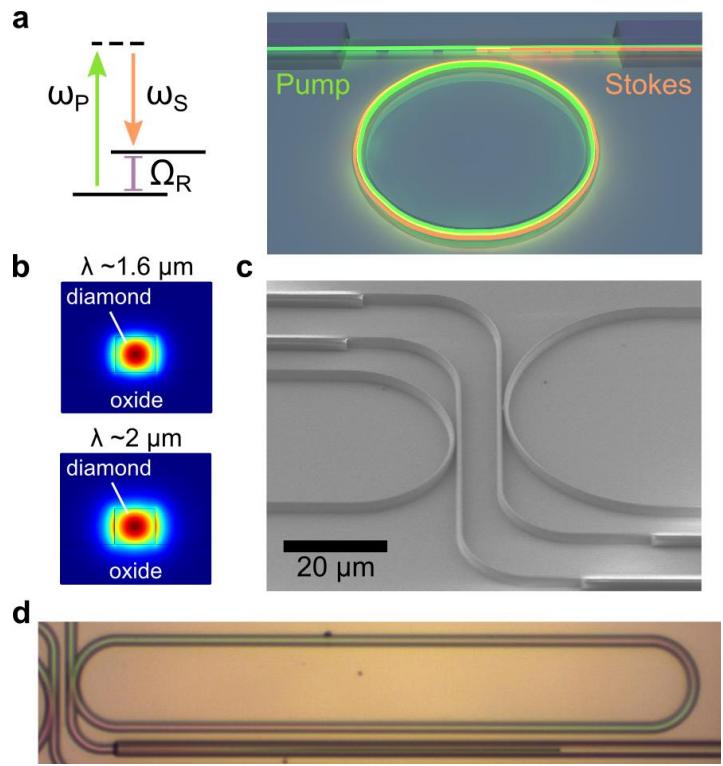

**Figure 1 | Diamond-microresonator based Raman laser design. a**, Energy level diagram of the Raman scattering process (left), wherein a high energy pump photon with frequency $\omega_P$ is scattered into a lower frequency Stokes photon, $\omega_S$, and an optical phonon, $\Omega_R$ (~40 THz in diamond). We pump with telecom lasers ($\lambda_P$ ~ 1.6 µm) corresponding to $\omega_P$ ~ 190 THz, resulting in a Stokes output at $\omega_S$ ~ 150 THz i.e. $\lambda_S$ ~ 2 µm. A schematic illustrating the device principle (right) shows a pump wave (green) entering a high-Q microcavity, where it enables Stokes lasing (yellow) via stimulated Raman scattering. **b**, Simulated mode profiles of diamond waveguides with width 800 nm and height 700 nm embedded in silica, at the pump ($\lambda_P$ ~1.6 µm, top) and Stokes $\lambda_S$ ~2 µm, bottom) wavelengths, showing good overlap. **c**, Scanning-electron-microscopy image of the nano-fabricated diamond racetrack resonators on an $SiO_2$-on-Si substrate before cladding with PECVD silica, showing the bus-waveguide-coupling region (gap ~ 500 nm) and transition to polymer (SU-8) waveguides for efficient coupling to lensed fibers. **d,** Optical micrograph of a diamond racetrack micro-resonator after a PECVD silica layer is deposited with path length ~600 µm and bending radius ~20 µm.

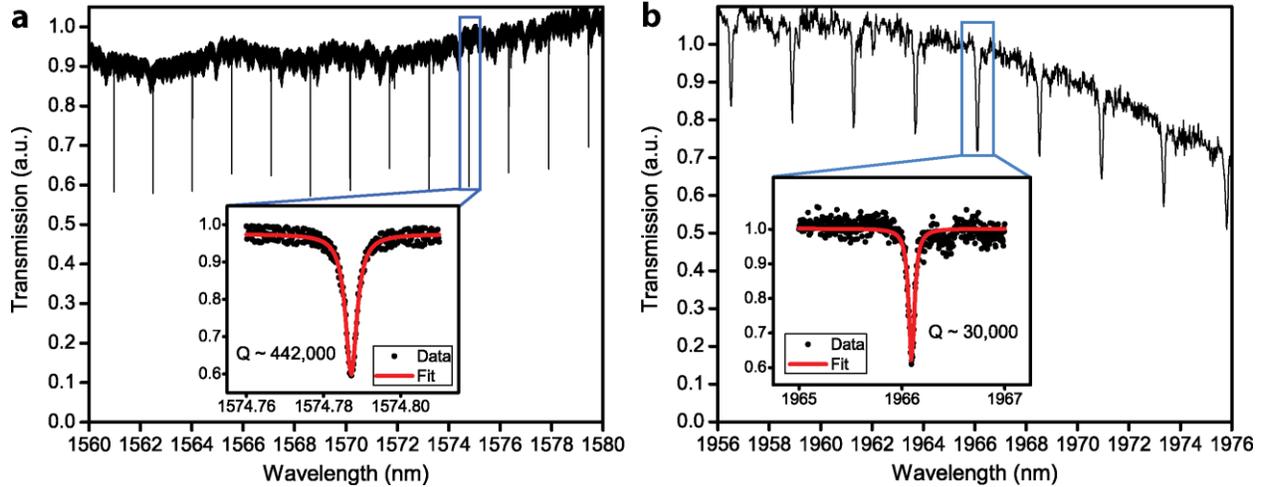

**Figure 2 | High-Q modes at pump and Stokes wavelengths. a**, Transmission spectrum of the diamond racetrack resonator at telecom (pump) wavelengths taken by sweeping a continuous-wave laser reveals high-Q transverse-electric (TE) modes with 30-40% extinction ratio (under-coupled resonances). The path length of the resonator is ~600 µm, corresponding to an FSR of ~1.5 nm (~180 GHz). Inset: A loaded Q of ~440,000 is inferred from the Lorentzian fit to the mode at ~1574.8 nm. **b**, , Transmission spectrum of the diamond resonator at the Stokes wavelength range near ~2-µm (~40 THz red-shifted from the pump) taken using a broadband super-continuum source again reveals high-Q TE modes with 30-40% extinction ratio (under-coupled resonances). Inset: A loaded Q of ~30,000 is inferred from the Lorentzian fit to the mode at ~1966 nm, although this may be limited by the resolution (~0.056 nm) of our optical spectrum analyzer.

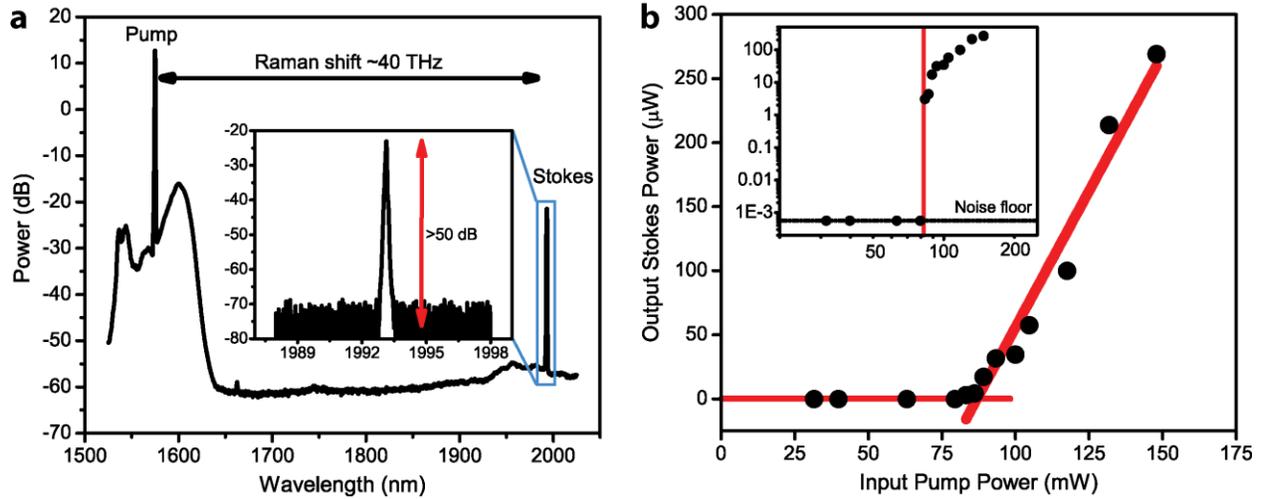

**Figure 3 | Observation of Raman lasing and threshold measurement. a**, Optical spectrum analyzer (OSA) signal when the pump is tuned into a resonance near ~1575 nm with ~100 mW power shows the emergence of the Raman line at the Stokes wavelength of ~1993 nm, ~40 THz red-shifted from the pump. Inset: A high-resolution scan zooming into the Stokes output reveals >50 dB sideband suppression ratio (>60 dB on-chip after correcting for out-coupling losses). **b**, Output Stokes power at ~1993 nm versus input pump power at ~1575 nm (both estimated in the bus-waveguide), displaying a clear threshold for Raman lasing at ~85 mW pump power. The external conversion slope efficiency is ~0.43%, corresponding to an internal quantum efficiency of ~12%. Inset: A log-log plot of the output Stokes power versus input pump power reveals a ~40 dB jump above the noise floor in the output at threshold.

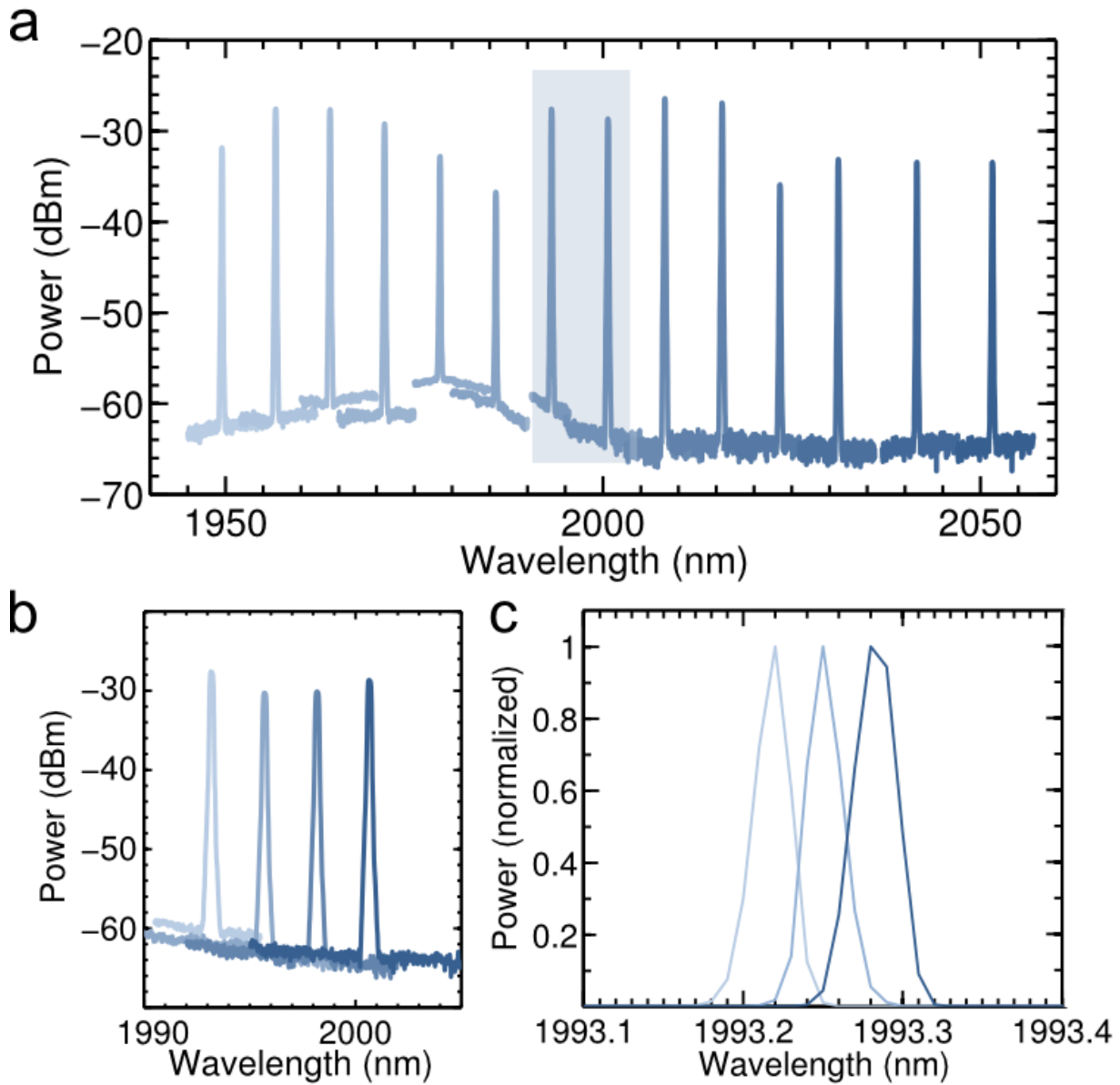

**Figure 4 | Discrete and continuous tuning of Raman laser output wavelength. a**, Discrete tuning of the Stokes wavelength over a range >100 nm (~7.5 THz or ~5% of the center frequency). The pump is tuned to 14 separate resonances, each spaced by 3xFSR (~550 GHz), and the Raman line is recorded with an optical spectrum analyzer (OSA) at each pump wavelength. **b**, Stokes output of adjacent modes. Here the pump is tuned to neighboring resonances (one FSR apart) within the highlighted region of Fig. 4a. The output modes are also spaced by an FSR or ~180 GHz. Thus, over 40 individual longitudinal modes can be accessed over the entire demonstrated tuning range. **c**, Mode-hop-free tuning of the Stokes wavelength over ~0.1 nm or ~7.5 GHz. The pump frequency is tuned within a thermally red-shifted resonance ('shark-fin' shape), thus tuning the output Stokes wavelength in a continuous fashion. The output power is normalized to the peak emission at each pump wavelength. The linewidth of the Stokes mode is limited by the minimum resolution of our OSA (~0.05 nm).